\documentclass[12pt,a4paper,twoside]{article}

\usepackage{amssymb,amsmath,bm,graphicx}

\title{Direct calculation of the probability of~pionium~ionization
in~the~target}
\author{M.\:V.~Zhabitsky\thanks{zhabitsk@nusun.jinr.ru}\protect\\
\textit{Joint Institute for Nuclear Research, Dubna, Russia}}
\date{Submitted to ``Physics of Atomic Nuclei''\thanks{\copyright 2008
Pleiades Publishing, Inc.}
(``Yadernaya Fizika''\thanks{\copyright 2008 MAIK Nauka/Interperiodica})}

\begin{document}

\maketitle

\begin{abstract}
  The goal of the DIRAC experiment at CERN
  is the lifetime measurement of pionium ($\pi^+\pi^-$~atom).
  Its lifetime is mainly defined by the charge-exchange process
  $\pi^+\pi^-\rightarrow\pi^0\pi^0$.
  Value of the lifetime in the ground state
  is predicted in the framework of the Chiral Perturbation
  Theory with high precision:
  $\tau_{1S}=(2.9\pm0.1)\cdot 10^{-15}$~s.
  The method used by DIRAC
  is based on analysis of $\pi^+\pi^-$-pairs spectra
  with small relative momenta
  in their center of mass system in order to find out signal
  from pionium ionization (break-up) in the target.
  Pioniums are produced in proton-nuclei collisions and
  have relativistic velocities ($\gamma>10$). 
  For fixed values of the pionium momentum and the target thickness
  the probability of pionium ionization in the target depends on
  its lifetime in a unique way,
  thus the pionium lifetime can be deduced from
  the experimentally defined probability of pionium ionization.
  Based on ionization cross-sections of pionium
  with target atoms
  we perform the first direct calculation of the pionium ionization probability
  in the target.
\end{abstract}

\section*{Introduction}
Pionium is the hydrogen-like Coulomb bound system
of two oppositely charged pions.
Its lifetime is determined
by strong $\pi^+\pi^-\rightarrow\pi^0\pi^0$ annihilation,
but it interacts mainly electromagnetically with the target atoms,
while propagating through the target material
with the relativistic velocity.
The DIRAC experiment at CERN~\cite{DIRACproposal}
detects $\pi^+\pi^-$-pairs with low relative momenta
in their center of mass system.
From the low-momentum part of the spectrum,
the probability of pionium ionization (break-up) in the target,
which is the probability for the pionium to be converted into an unbound
$\pi^+\pi^-$~pair on the exit of the target,
is determined.
If the dependence of the pionium ionization probability
in the target as a function of its lifetime is established,
this lifetime measurement may then be confronted
with the value predicted in the framework
of the Chiral Perturbation Theory~\cite{GasserPRD2001}.
The measurement by DIRAC will test the standard picture
of spontaneous chiral symmetry breaking in QCD,
providing the experimental evidence in favour of or against the existence
of a large quark-antiquark condensate in the QCD vacuum~\cite{Knecht}.

The main thrust of the DIRAC technique
is that the dependence of the pionium ionization probability
in the target as a function of its lifetime
can be established with sufficient precision.
Despite the progress in calculations
of ionization and excitation cross-sections
from different pionium states~\cite{baselCS},
it proved to be difficult to solve the infinite system
of differential (kinetic) equations, which describes the evolution
of a pionium propagating through the target.
In the original studies~\cite{Afan1996,Cibran2003}
probabilities for the pionium to leave the target in a bound state
or annihilate
were calculated by taking into account only low-lying atomic states.
The contribution of highly-excited states was estimated
by extrapolation and the sought-after probability
of pionium ionization in the target was obtained
indirectly by means of unitarity.
The precision of the indirect calculation
was challenged by the fact, that pionium excitation
to ever higher lying bound states constitutes a major branch
in the evolution of pionium.

In this work we solve the system of kinetic equations
by taking into account dynamics of highly excited pionium bound states.
Within our approach
the pionium ionization probability in the target is calculated directly,
based on ionization cross-sections.
The formalism of pionium dynamics,
described by a set of the probabilistic kinetic equations,
is reminded in Section~\ref{sect:max}.
Section~\ref{sect:min} is devoted to the direct
calculation of the ionization probability.

\section{Dynamics of a pionium in the target}
\label{sect:max}
The pionic atoms can be created in inelastic proton-nuclei collisions
with the probability given by~\cite{Nem85}
\[
 \frac{d\sigma_A}{d\vec P} = 
 (2\pi)^3\left|\Psi(0)\right|^2
 \frac{E}{M}
 \left.\frac{d\sigma^0_s}{d\vec p_+ d\vec p_-}
  \right|_{\vec p_+\approx \vec p_-},
\]
where $\displaystyle \frac{d\sigma^0_s}{d\vec p_+ d\vec p_-}$
is the double inclusive cross section for production of $\pi^+\pi^-$~pairs
without interaction in the final state
with both pions produced either directly in hadronic processes
or through short-lived resonances.
$\vec P$, $E$ and $M$ are the pionium momentum, energy
and mass in the laboratory frame, respectively,
$\vec p_-$ and $\vec p_+$ are the momenta of the charged pions.
Production of the pionium atoms
with the angular momentum \mbox{$l>0$} is suppressed.
The strong interaction between the two pions, which forms the atom,
significantly modifies $|\Psi_{n0}(0)|$
in comparison to the pure Coulomb wave functions~\cite{Kuraev98}.
However, only ratios between the squares
of the wave functions modulus at the origin
enter to the following calculations.
For these ratios the modification due to the strong interaction
is almost cancelled.
The probability of pionium production
depends on the principal quantum number $n$ as~\cite{Lednicky}
\[
  (1+\Delta_n)\left|\Psi^C_{n0}(0)\right|^2
 \propto (1+\Delta_n)\frac{1}{n^3}
 \approx \left(1+\frac{0.003}{n}\right)\frac{1}{n^3},
\]
where $\Psi^C_{n0}(0)$ is the pure Coulomb wave function
of the $\pi^+\pi^-$ atom at zero distance
and the correction factor $(1+\Delta_n)$
takes into account the effect of finite-size
of the pion production region and the two-pion
strong interaction in the final state.
It was found that this correction
shifts the probability of ionization
on per mille level,
therefore we will use pure Coulomb wave functions hereafter.
For them only $nS$ states are non-zero at the origin.
If we normalize probability of atom production to unity
then the probability of atom production in the $|nlm\rangle$ state reads
\begin{gather}
 p_{nlm}(0) = \frac{\left|\Psi^C_{nlm}(0)\right|^2}
                   {\sum_{k=1}^\infty \left|\Psi^C_{kS}(0)\right|^2}=
 \frac{\delta_{l0}}{n^3\sum_{k=1}^\infty 1/k^3}=
 \frac{\delta_{l0}}{n^3\zeta(3)}, \label{eq:pionium_prod}\\
p_{100}(0)=0.832,\quad p_{200}(0)=0.104,\quad p_{300}(0)=0.031.\nonumber
\end{gather}

After production atom can either annihilate
(mainly through $\pi^+\pi^-\rightarrow\pi^0\pi^0$ process\footnote{Another
annihilation channel $\pi^+\pi^-\rightarrow2\gamma$
amounts only about $0.3$\%~\cite{Uretsky61,Hammer99}.})
or electromagnetically interact with target atoms.

The partial decay width
of the pionium in $1S$ state into $2\pi^0$ is
\cite{GasserPRD2001}:
\[
 \Gamma_{2\pi^0}=\frac{2}{9}\alpha^3
 \sqrt{m_{\pi^+}^2-m_{\pi^0}^2-\frac{1}{4}m_{\pi^+}^2\alpha^2}
 (a^0_0-a^2_0)^2 m_{\pi^+}^2 (1+\delta_\Gamma),\quad
 \delta_\Gamma=(5.8\pm 1.2)\times 10^{-2}.
\]
The $(a^0_0-a^2_0)$ difference of the pion-pion S-wave scattering lengths
with isospin $0$ and $2$
has been calculated~\cite{ColangeloNPB603} within the framework
of the standard chiral perturbation theory (ChPT)~\cite{ChPT}
\[
 a^0_0-a^2_0 = (0.265\pm 0.004) m_{\pi^+}^{-1}.
\]
This difference leads to the predicted value of the pionium lifetime
in the ground state
\begin{equation}
 \tau_{1S}=(2.9\pm 0.1)\cdot 10^{-15}\:s.
 \label{eq:lifetime}
\end{equation}
Lifetime of the atom is inversely proportional to the square
of its wave function modulus at the origin~\cite{Bilenky69}, so
in $nS$ states it reads
$\tau_{nS}=
\displaystyle\tau_{1S}\frac{|\Psi_{1S}(0)|^2}{|\Psi_{nS}(0)|^2}\approx
\displaystyle\tau_{1S}\frac{|\Psi^C_{1S}(0)|^2}{|\Psi^C_{nS}(0)|^2}=
\tau_{1S}n^3$.
Therefore the probability for a pionium
with the laboratory momentum~$P$
to annihilate per unit length is
\[
W_{nlm}^\text{anh} = \frac{1}{\lambda_{nlm}^\text{anh}}=
  \begin{cases}
           \displaystyle\frac{1}{\gamma \beta \tau_{nS}}=
           \frac{2m_\pi}{P \tau_{1S}n^3} & 
              \text{in $nS$ states,}\\
            0 & \text{in other states.}
  \end{cases}
\]

While crossing the target
a pionium electromagnetically interacts
with target atoms.
As a result, $\pi^+\pi^-$~atom can be either ionized
or transit from the initial bound state $|n_i l_i m_i\rangle$
to another bound state $|n_f l_f m_f\rangle$
(excitation/de-excitation).
Hereafter we will denote initial and final bound states
as $|i\rangle$ and $|f\rangle$, respectively.

The probability of ionization per unit length
from the state $|i\rangle$ is given by
\[
 W_i^\text{ion} = 
 \frac{1}{\lambda_i^\text{ion}} =
 \frac{\rho N_A}{A}\sigma_i^\text{ion},
\]
where $\rho$ is the target density,
$A$ is its atomic weight,
$N_A$ is the Avogadro constant
and $\sigma_i^\text{ion}$ is the ionization cross section.

The probability of a pionium excitation per unit length
from the state $|i\rangle$ to the final state $|f\rangle$ is given by
\[
 W_i^f =
 \frac{1}{\lambda_i^f} =
 \frac{\rho N_A}{A}\sigma_i^f,
\]
where $\sigma_i^f$ is
the discrete (bound-bound) transition cross section.

The total inelastic cross section gives the probability
of an atom to undergo an inelastic electromagnetic interaction
\begin{equation}
 {\sigma_i^\text{tot} = \sum\limits_f \sigma_i^f + \sigma_i^\text{ion}.}
 \label{eq:sigmatot}
\end{equation}
Total cross sections can be calculated
owing to the completeness of the eigenstates
of the Coulomb Hamiltonian.

Total and transition cross sections for any bound states
were initially calculated in the Born approximation
with the static potential of target atoms~\cite{Afan1996}.
Later a more accurate set of cross sections was derived
which takes into account relativistic effects,
multiple photon exchange
and target excitations~\cite{baselCS}.
Moreover, in the latter work authors calculated
ionization cross sections 
for any initial bound state with $n\leqslant 8$ 
which provides the possibility
to perform direct calculation of the pionium ionization probability
in the target, which is the subject of the present work.
From general formulas for transition cross sections
in both sets~\cite{Afan1996,baselCS}
one can derive that for any states $|i\rangle$  and $|f\rangle$
$\sigma_i^f=\sigma_f^i$.
Comparison between different sets of cross sections
was performed~\cite{Cibran2003},
where authors found that 
uncertainties in most precise sets of cross-sections
for Ni target
will cause only 1\% uncertainty in the pionium lifetime.
Uncertainty due to the accuracy of cross sections
is expected to dominate precision of the ionization probability
dependence on the pionium lifetime.

The dynamics of the pionium interaction with target atoms
is supposed to be described by a set of kinetic equations~\cite{Afan1996}
using the probabilities~$p_i(s)$ to find the $\pi^+\pi^-$ atom
in the definite quantum state~$|i\rangle$
at a distance~$s$ from the production point.
This approach ignores any interference between different
pionium bound states.
For low $n$
most of interference effects are suppressed
at typical pionium momenta in DIRAC ($3\div 8\:\text{GeV}/c$)
as the mean free path between pionium inelastic interactions
is usually longer than the formation time of atomic system
multiplied by its velocity.
Nevertheless even for low $n$
some interference effects can take place
due to the accidental degeneracy of energy levels
of hydrogen-like atoms.
This problem was considered
in the framework of the density matrix formalism~\cite{densmatr04}.
It was found that the interference between quantum states with small $n$
does not change the result based on a set of probabilistic kinetic equations
(their difference is less than per mille).

Eigenstates of the Coulomb Hamiltonian
form a countable set of discrete levels.
For numerical calculations we will take into account
only levels with
a principal quantum number ${n\leqslant n_\text{max}}$.
For a given principal quantum number $n$ there are $n^2$
states $|nlm\rangle$ with different orbital and magnetic quantum numbers.
We will denote the total number of discrete bound states
taken for the calculations as $N$.
To make the system complete we introduce 
a cross section ${\sigma^u_i}$
which stands for the sum of transitions from state $|i\rangle$ to any
discrete state with a principal quantum number
$n_f>n_\text{max}$:
\[
  {\sigma^u_i = 
 \sum_{f:n_f>n_\text{max}} \sigma_i^f =
 \sigma^{tot}_i - \sigma^\text{ion}_i -
 \sum_{f:n_f\leqslant n_\text{max}}\sigma^f_i.}
\]
It is straightforward to write
the probability of pionium production
in all bound states above ${n_\text{max}}$: 
\begin{equation}
  p_u(0) = 1 - \sum_{i: n_i\leqslant n_\text{max}} p_i(0).
  \label{eq:p_u_initial}
\end{equation}

Finally we will write the system of kinetic equations
in the matrix form
\begin{equation}
\frac{d}{ds}
\begin{pmatrix}
p_1\\
p_2\\
\ldots\\
p_N\\
p_u\\
p_\text{ion}\\
p_\text{anh}
\end{pmatrix}=
\begin{pmatrix}
W^1_1  & W^1_2  & \ldots & W^1_N  & 0 & 0 & 0 \\
W^2_1  & W^2_2  & \ldots & W^2_N  & 0 & 0 & 0 \\[-5pt]
\vdots & \vdots & \ddots & \vdots & 0 & 0 & 0 \\
W^N_1  & W^N_2  & \ldots & W^N_N  & 0 & 0 & 0 \\
W^u_1  & W^u_2  & \ldots & W^u_N  & 0 & 0 & 0 \\
W^\text{ion}_1 & W^\text{ion}_2 & \ldots & W^\text{ion}_N & 0 & 0 & 0 \\
W^\text{anh}_1 & W^\text{anh}_2 & \ldots & W^\text{anh}_N & 0 & 0 & 0
\end{pmatrix}
\begin{pmatrix}
p_1\\
p_2\\
\ldots\\
p_N\\
p_u\\
p_\text{ion}\\
p_\text{anh}
\end{pmatrix}.
\label{eq:system0}
\end{equation}
Diagonal terms describe the total decrease of the level population
\[
 W^i_i = -\frac{\rho N_A}{A}\sigma^\text{tot}_i - W_i^\text{anh}.
\]
System~(\ref{eq:system0}) is a system of linear homogeneous
differential equations of order~1
with constant coefficients.
The rank of the matrix is $N$,
with three low lines being a linear combination of the first $N$~lines.
It is exactly solvable
\begin{equation}
 p_i(s) = \sum\limits_k c_k
  \alpha_i^{\left(k\right)} e^{\lambda_k s},
 \label{eq:eigen_sol}
\end{equation}
where $\lambda_1,\ldots,\lambda_N$ are eigenvalues
and $\alpha^{\left(k\right)}$ their corresponding eigenvectors.
Symmetry of the upper left $N$-by-$N$ corner
guaranties that all its eigenvalues are real~\cite{linalg}.
Coefficients ${c_k}$ are fixed from initial
conditions (\ref{eq:pionium_prod}), (\ref{eq:p_u_initial}).
The probability of pionium ionization
at the distance~$s$ after the production point
is expressed through the solution~(\ref{eq:eigen_sol})
\[
 p_\text{ion}(s) = \sum\limits_k
  \frac{c_k}{\lambda_k}
  \left( e^{\lambda_k s} - 1 \right)
  \sum\limits_i W_i^\text{ion} \alpha_i^{\left(k\right)}.
\]

DIRAC uses very thin targets (their nuclear efficiency
is less than $10^{-3}$), therefore atoms are produced
nearly uniformly over the target thickness~$s_0$.
Hence the probability for a pionium to leave the target
in the state $|i\rangle$ reads
\[
  P_i(s_0) =  \frac{1}{s_0} \int_0^{s_0} p_i(s) ds =
  \sum\limits_k c_k
  \alpha_i^{\left(k\right)} \frac{1}{\lambda_k s_0}
  \left( e^{\lambda_k s_0} -1 \right),
\]
while the probability of ionization on the exit of the target is
\begin{equation}
 P_\text{ion}(s_0) = \sum\limits_k
  \frac{c_k}{\lambda_k}
   \left(
    \frac{1}{\lambda_k s_0} \left(e^{\lambda_k s_0} - 1 \right) -1
   \right) \sum\limits_i W_i^\text{ion} \alpha_i^{\left(k\right)}.
 \label{eq:ion_sol_av}
\end{equation}
Expressions for the probability
of annihilation $P_\text{anh}$ and
for the probability $P_{u}$
to reach any excited state with $n>n_\text{max}$ on the exit of the target
have the same form
as (\ref{eq:ion_sol_av}) if one substitutes $W_i^\text{ion}$
with $W_i^\text{anh}$ and $W_i^u$ respectively.

\begin{table}
\caption{Results for different probabilities
on the exit of the target
as a function of $n_\text{max}$
if the dynamics of highly excited states (with $n>n_\text{max}$)
is not taken into account}
\centering
\bigskip
\begin{tabular}{|c|l|l|l|l|}
\hline
\rule{0pt}{14pt}%
${n_\text{max}}$ &
 ${P_\text{dsc}^A}$ & 
               ${P_\text{anh}^A}$ &
                         ${P_\text{ion}^A}$ &
                                   ${P_\text{u}^A}$ \\[2pt] \hline
\rule{0pt}{14pt}%
5 & 0.0928 & 0.4407 & 0.2438 & 0.2227\\
6 & 0.0937 & 0.4407 & 0.2586 & 0.2070\\
7 & 0.0943 & 0.4407 & 0.2715 & 0.1935\\
8 & 0.0947 & 0.4407 & 0.2828 & 0.1817\\[1mm]\hline
\end{tabular}
 \label{tab:noevol}
\end{table}

In Table~\ref{tab:noevol} we illustrate this solution
as a function of $n_\text{max}$
for a pionium atom produced in $95\:\mu$m thick Ni~target
with the momentum $P=4.6\:\text{GeV}/c$,
corresponding to the average laboratory momentum of produced pioniums
in the kinematic range of the DIRAC experiment.
Eigenvalues were numerically found by functions
from the LAPACK~\cite{LAPACK} package.
System of equations~(\ref{eq:system0}) is constructed in a way
that as soon as an atom reaches a state with $n>n_\text{max}$
it effectively quits from calculations and is kept intact,
though in reality it is expected to undergo further electromagnetic
interactions, e.\:g. it can be ionized or
de-excited to the low-lying states.
Therefore, $P_\text{dsc}^A$ is at least
the probability for a pionium to leave the target
in any bound state with the principal quantum number
$n\leqslant n_\text{max}$.
${P_\text{u}^A}$ is the probability for atoms
to reach states with ${n>n_\text{max}}$,
which amounts to about 20\%.
This numerical value is in agreement
with the earlier calculations~\cite{Cibran2003} (Fig.\:3(d)).
Numerical precision of the above solution can be estimated
from the inequality
\[
 \left|1 - P_\text{dsc}^A-P_\text{anh}^A-P_\text{ion}^A-P_\text{u}^A\right|
 < 1\cdot 10^{-12},
\]
thus round-off errors do not affect the result.

As an atom transits to a state $n_f>n_i$ its
effective radius of electromagnetic interactions grows
and its characteristic ionization length is getting shorter
\[
\lambda^\text{ion}_{|n>n_\text{max},lm\rangle} <
  \lambda^\text{ion}_{|n_\text{max}=8,lm\rangle} \approx 2\:\mu\text{m}.
\]
The target used in DIRAC is $95\mu$m thick, therefore
highly excited atoms have a chance to leave the target in a bound state
only if they were created close to the exit of the target.
Otherwise these highly excited atoms will be ionized.
This allows us to set the range for the ionization probability
in the target:
\begin{equation}
 0.2828 = P^A_\text{ion} < P_\text{ion} <  P^A_\text{ion} + P^A_u= 0.4645.
 \label{eq:PbrA}
\end{equation}
Here the upper bound corresponds to the case
when all highly excited atoms are ionized,
while the lower bound ${P_\text{ion}^A}$ is at least probability
of ionization from states
with ${n\leqslant n_\text{max}}$.

From Table~\ref{tab:noevol} one can conclude that
above upper and lower bounds converge slowly
with increase of ${n_\text{max}}$
and in this way
it would be difficult to increase ${n_\text{max}}$ in order to achieve
precision required by DIRAC (per cent level).

\section{Evolution of highly-excited states}
\label{sect:min}
\begin{figure}
 \centering
 \includegraphics[width=0.7\textwidth]{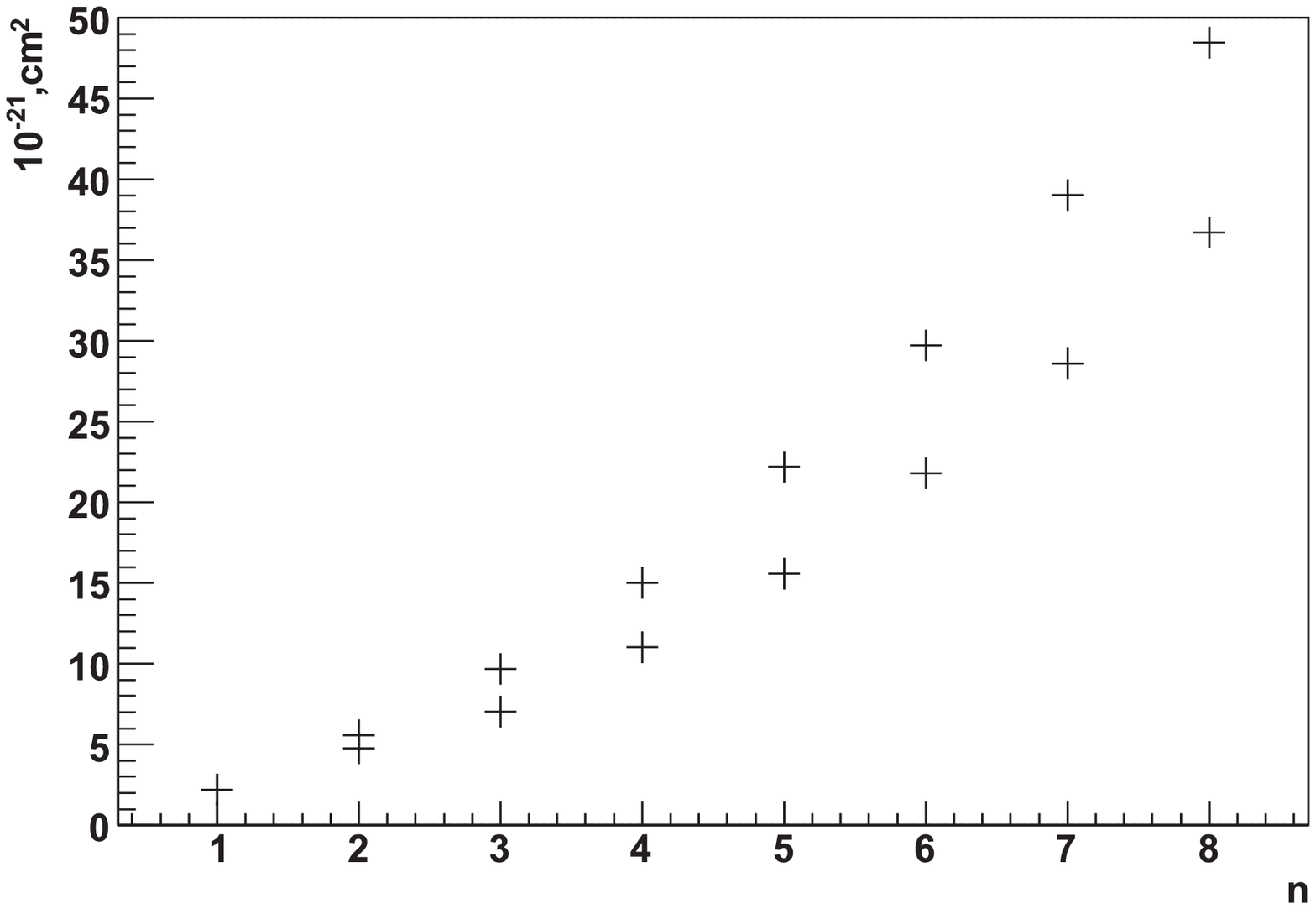}
 \caption{The minimal $\displaystyle\left(\min_{l^{\prime}m^{\prime}} \left\{
      \sigma^\text{ion}_{|nl^\prime m^\prime\rangle}\right\}\right)$
      and the maximal $\displaystyle\left(\max_{l^{\prime}m^{\prime}} \left\{
      \sigma^\text{ion}_{|nl^\prime m^\prime\rangle}\right\}\right)$
      pionium ionization cross sections on the Ni atom
      for different initial principal quantum numbers~$n$ of the pionium
      (according to \cite{baselCS}).}
 \label{fig:sigmaion_range}
\end{figure}

Rather than trying to solve the system~(\ref{eq:system0}) directly
we will modify it in order
to get the \textit{lower} bound of the ionization probability
by taking into account dynamics of highly excited states
with $n>n_\text{max}$:
\begin{equation}
\frac{d}{ds}
\begin{pmatrix}
p_1\\
p_2\\
\ldots\\
p_N\\
p_u\\
p_\text{ion}\\
p_\text{anh}
\end{pmatrix}=
\begin{pmatrix}
W^1_1  & W^1_2  & \ldots & W^1_N  & W^1_u & 0 & 0 \\
W^2_1  & W^2_2  & \ldots & W^2_N  & W^2_u & 0 & 0 \\[-5pt]
\vdots & \vdots & \ddots & \vdots & \vdots & 0 & 0 \\
W^N_1  & W^N_2  & \ldots & W^N_N  & W^N_u & 0 & 0 \\
W^u_1  & W^u_2  & \ldots & W^u_N  & W^u_u & 0 & 0 \\
W^\text{ion}_1 & W^\text{ion}_2 & \ldots & W^\text{ion}_N & W^\text{ion}_u & 0 & 0 \\
W^\text{anh}_1 & W^\text{anh}_2 & \ldots & W^\text{anh}_N & W^\text{anh}_u & 0 & 0
\end{pmatrix}
\begin{pmatrix}
p_1\\
p_2\\
\ldots\\
p_N\\
p_u\\
p_\text{ion}\\
p_\text{anh}
\end{pmatrix}.
\label{eq:system_min}
\end{equation}
Here
\[
 W_u^\text{ion} = 
 \frac{\rho N_A}{A} \sigma^\text{ion}_{n_\text{max}+1,\min},\quad
 \sigma^\text{ion}_{n_\text{max}+1,\min} =
  \min_{l^{\prime}m^{\prime}} \left\{
      \sigma^\text{ion}_{|n_\text{max}+1, l^\prime m^\prime\rangle}
  \right\}
\]
is the lower bound of the probability of ionization
per unit length
from any state with $n>n_\text{max}$,
because the ionization cross section tends to grow
with increasing of the principal quantum number~$n$
due to the corresponding expansion of the atomic radius.
The minimal and maximal values of the ionization
cross-section for different principal quantum numbers
are drawn in Fig.~\ref{fig:sigmaion_range}. 
To find the lower bound of the ionization probability,
further we require all probabilities of ionization
per unit length from any state $|nlm\rangle$
do not exceed $W_u^\text{ion}$:
\[
  W_{|nlm\rangle, \min}^\text{ion} = 
 \frac{\rho N_A}{A} \sigma^\text{ion}_{|nlm\rangle,\min},\quad
 \sigma^\text{ion}_{|nlm\rangle,\min}
 = \min\left\{ \sigma^\text{ion}_{|nlm\rangle},\ 
              \sigma^\text{ion}_{n_\text{max}+1,\min}
        \right\}.
\]
The diagonal term $W_i^i$, which describes the level de-population,
is changed accordingly to fulfill~(\ref{eq:sigmatot}).

Upper bound of
the probability of de-excitation per unit length
from all states with $n>n_\text{max}$
to a state $|f\rangle$ with $n_f\leqslant n_\text{max}$
is obtained from the following inequality
\[
 \sum_{i:n_i>n_\text{max}} W_i^f p_i < 
 \sum_{i:n_i>n_\text{max}} W_i^f \cdot \sum_{i:n_i>n_\text{max}}p_i =
 W_u^f p_u =
 W_f^u p_u.
\]
Finally the diagonal term,
which describes de-population of discrete states
with $n>n_\text{max}$, is
\[
 W^u_u = -\frac{\rho N_A}{A}\sigma_{n_\text{max}+1,\min}^\text{ion}
  - W_u^\text{anh}
  - \sum_{f:n_f\leqslant n_\text{max}} W_u^f,
\]
where $W_u^\text{anh}=\displaystyle\frac{2m_\pi}{P \tau_{1S}(n_\text{max}+1)^3}$
is the upper bound of
the probability for a pionium to annihilate
from any state with $n>n_\text{max}$ per unit length.
The rank of the new system is $N{+}1$.
The system~(\ref{eq:system_min}) is constructed in the way
that ionization is \textit{underestimated} and
all competitive processes including de-excitation from high ${n}$ states
(thus transitions to bound states with even lower ionization)
are \textit{overestimated},
therefore the solution is the \textit{mathematical lower bound}
of the probability of ionization.
Numerical results for different $n_\text{max}$
are presented in Tab.~\ref{eq:resultsB}.

\begin{table}
\caption{Numerical solution of the system~(\ref{eq:system_min})
 for the lower bound of the probability of pionium ionization in the target
 as a function of $n_\text{max}$}
\bigskip
\centering
\begin{tabular}{|c|l|l|l|l|}
\hline
\rule{0pt}{14pt}%
${n_\text{max}}$ &
 ${P_\text{dsc}^B}$ & 
               ${P_\text{anh}^B}$ &
                         ${P_\text{ion}^B}$ &
                                   ${P_\text{u}^B}$ \\[2pt] \hline
\rule{0pt}{14pt}%
5 & 0.1109 & 0.4416 & 0.4468 & 0.00067\\
6 & 0.1068 & 0.4411 & 0.4517 & 0.00030\\
7 & 0.1041 & 0.4409 & 0.4548 & 0.00015\\
8 & 0.1023 & 0.4408 & 0.4567 & 0.00008\\[1mm]\hline
\end{tabular}
\label{eq:resultsB}
\end{table}

\begin{figure}
 \centering
 \includegraphics[width=0.7\textwidth]{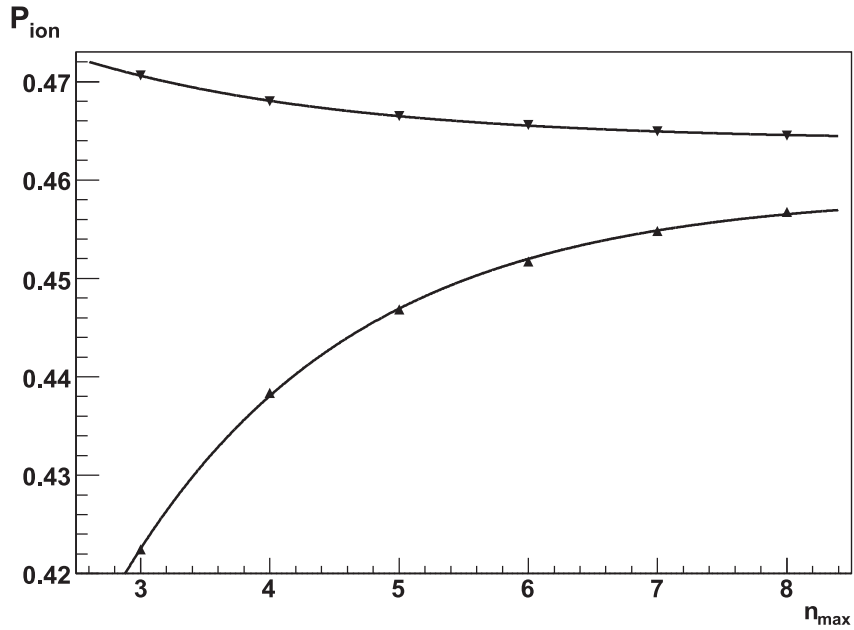}
 \caption{Upper and lower bounds of the probability of pionium ionization
          in the target $P_\text{ion}$
          as a function of $n_\text{max}$,
          fitted by $a e^{\alpha n_\text{max}}+c$ functions to guide the eye
          ($a$, $\alpha$ and $c$ are free parameters).}
 \label{fig:Pbr_min_max}
\end{figure}

Upper~(\ref{eq:PbrA}) and lower bounds effectively squeeze the solution
(Fig.~\ref{fig:Pbr_min_max}),
for $n_\text{max}=8$ they are
\begin{equation}
  0.4567 = P^B_\text{ion} <
  P_\text{ion} <  P^A_\text{ion} + P^A_u = 0.4645,\quad
  \frac{P_\text{ion}^\text{max}-P_\text{ion}^\text{min}}
       {2P_\text{ion}} \approx 0.8\cdot 10^{-2}.
  \label{eq:Pion_final}
\end{equation}
Precision of the above calculated value of the pionium ionization
probability in the target is comparable to the $\sim 1$\% uncertainties
in the $P_\text{ion}$ value due to precision
of the corresponding electromagnetic cross-sections~\cite{Cibran2003}.

\begin{figure}
 \centering
 \includegraphics[width=0.7\textwidth]{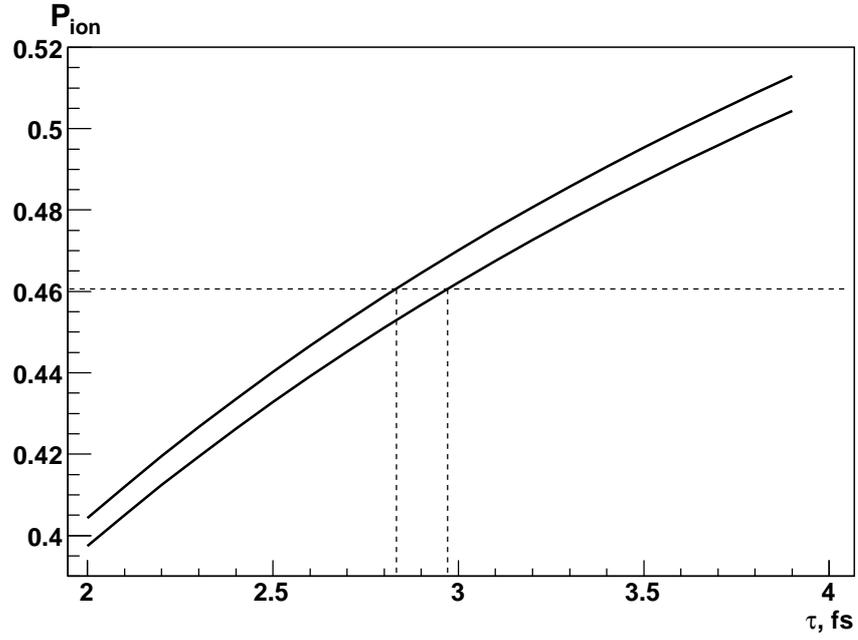}
 \caption{Upper and lower bounds of the pionium ionization probability
          in the target $P_\text{ion}$
          as a function of the pionium lifetime in the ground state.}
 \label{fig:Pbr_min_max_tau}
\end{figure}

Upper and lower bounds of the probability of pionium ionization in the target
as a function of its lifetime in the ground state
are shown in Fig.~\ref{fig:Pbr_min_max_tau}.
The corresponding uncertainties are shown as dashed lines
around the value of the pionium lifetime~(\ref{eq:lifetime})
predicted by theory.
The DIRAC collaboration reported the measured value of
$P_\text{ion}=0.452^{+0.025}_{-0.039}$~\cite{DIRAC2005},
based on part of the collected data.
While further analysis will reduce uncertainties of the experimental result,
uncertainties of the solution~(\ref{eq:Pion_final})
are expected to be within precision, required by DIRAC.
Range can be further shrunk by extrapolation
as shown in Fig.~\ref{fig:Pbr_min_max}.

\begin{figure}
 \centering
 \includegraphics[width=0.7\textwidth]{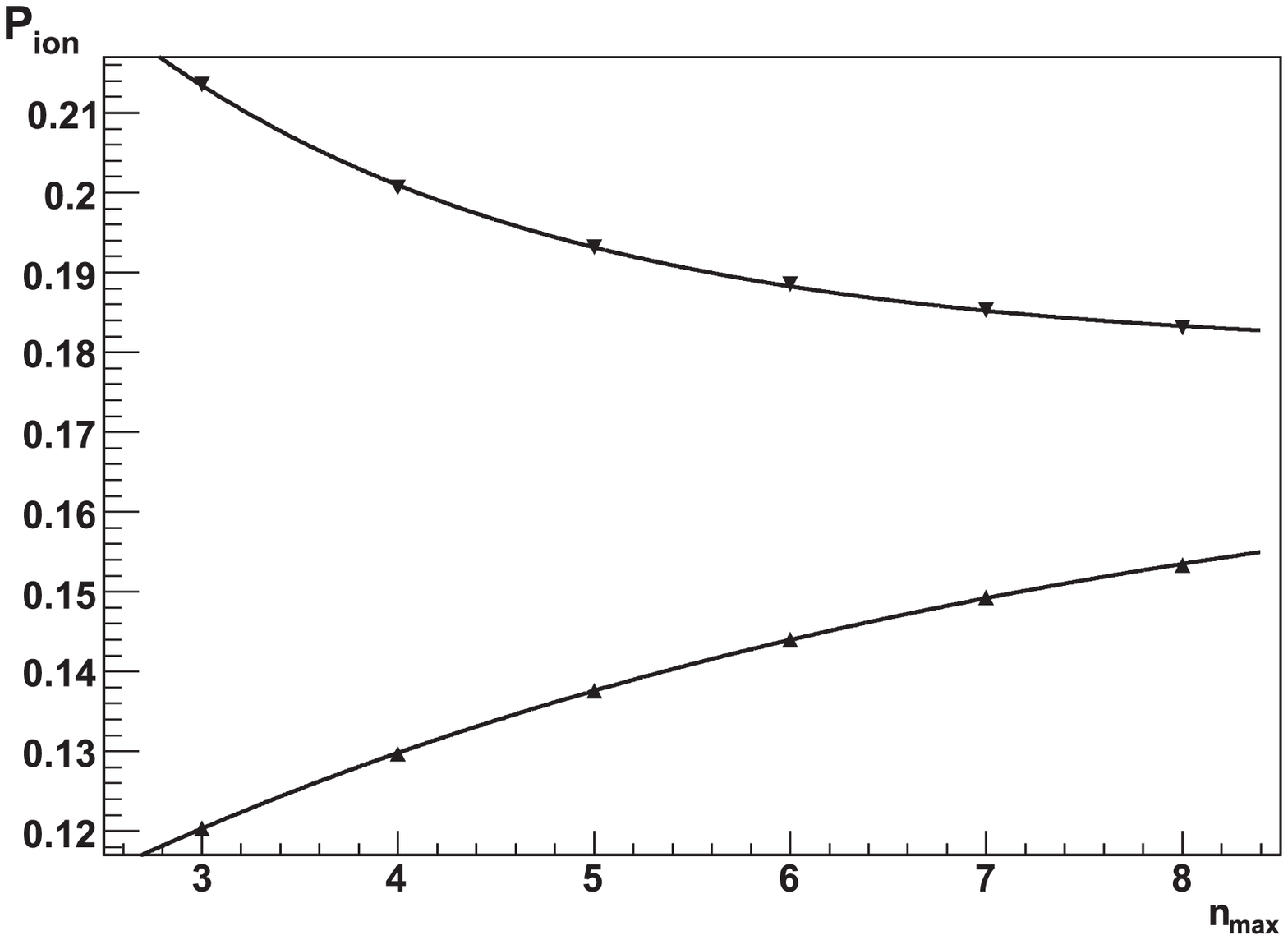}
 \caption{Upper and lower bounds of the pionium ionization probability
          in the target $P_\text{ion}$
          as a function of $n_\text{max}$
          for $10\:\mu$m thick Ni target.}
 \label{fig:Pbr_min_max_thin}
\end{figure}

We have to emphasize that
upper and lower bounds of the ionization probability
squeeze the solution with required precision
due to the fact,
that for atoms with a principal quantum number $n>8$
a characteristic ionization length
is less than $2\:\mu$m,
which is much shorter than the target thickness of $95\:\mu$m.
If one selects very thin target
(e.g. $10\:\mu$m thick Ni)
then the upper and lower bounds will show
worse convergence (Fig.~\ref{fig:Pbr_min_max_thin}).

\section*{Conclusions}

We derived a mathematical approach to solve a system
of kinetic equations, which describes evolution
of relativistic $\pi^+\pi^-$~atoms propagating through the target.
In this approach we reduce the system of kinetic equations,
which formally contains infinite number of equations,
to the finite set of equations, which is solved exactly.
The solution represents
lower and upper bounds for the probability of pionium ionization
in the target.
These lower and upper bounds effectively squeeze 
the solution to the value of the probability of ionization
with 1\% precision, which is within requirements of the DIRAC experiment.
Thus the first direct (based on ionization cross sections)
calculation of the probability of ionization has been performed.
We confirm that the contribution of highly-excited states
(with the principal quantum number $n>8$)
to the probability of ionization is significant ($>1/3$).

\bigskip
Author would like to thank
L.~Afanasyev, L.~Nemenov, A.~Tarasov and V.~Yazkov
for many discussions about the problem.

\newpage

\begin{thebibliography}{99}

\bibitem{DIRACproposal}
B.~Adeva \textit{et al.}, DIRAC proposal, CERN/SPSLC 95-1, SPSLC/P 284
(1995).

\bibitem{GasserPRD2001}
J.~Gasser, V.\:E.~Lyubovitskij, A.~Rusetsky and A.~Gall,
Phys. Rev.~D \textbf{64}, 016008 (2001).

\bibitem{Knecht}
M.~Knecht, B.~Moussallam, J.~Stern and N.\:H.~Fuchs,
Nucl. Phys. B \textbf{457}, 513 (1995).

\bibitem{baselCS}
Z.~Halabuka, T.~Heim, K.~Hencken, D.~Trautmann and R.\:D.~Viollier,
Nucl. Phys.~B \textbf{554}, 86 (1999);\\
T.~Heim, K.~Hencken, D.~Trautmann and G.~Baur,
J. Phys.~B: At. Mol. Opt. Phys. \textbf{33}, 3583 (2000);\\
T.~Heim, K.~Hencken, D.~Trautmann and G.~Baur,
J. Phys.~B: At. Mol. Opt. Phys. \textbf{34}, 3763 (2001).

\bibitem{Afan1996}
L.\:G.~Afanasyev and A.\:V.~Tarasov,
Yad. Fiz. \textbf{59}, 2212 (1996);
Phys. Atom. Nucl. \textbf{59}, 2130 (1996).

\bibitem{Cibran2003}
C.~Santamarina, M.~Schumann, L.\:G.~Afanasyev and T.~Heim,
J. Phys.~B: At. Mol. Opt. Phys. \textbf{36}, 4273 (2003).

\bibitem{Nem85}
L.\:L.~Nemenov, Yad. Fiz. \textbf{41}, 980 (1985);
Sov. J. Nucl. Phys. \textbf{41}, 629 (1985).

\bibitem{Kuraev98}
E.\:A.~Kuraev, Yad. Fiz. \textbf{61}, 378 (1998);
Phys. At. Nucl. \textbf{61}, 325 (1998).

\bibitem{Lednicky}
R.~Lednick\'y,
DIRAC internal note, 2005--18; arXiv:nucl-th/0501065.

\bibitem{Uretsky61}
J.\:L.~Uretsky and T.\:R.~Palfrey,
Phys. Rev. \textbf{121}, 1798 (1961).

\bibitem{Hammer99}
H.-W.~Hammer and J.\:N.~Ng,
Eur. Phys. J.~A \textbf{6}, 115 (1999).

\bibitem{ColangeloNPB603}
G.~Colangelo, J.~Gasser and H.~Leutwyler,
Nucl. Phys.~B \textbf{603}, 125 (2001).

\bibitem{ChPT}
J.~Gasser and H.~Leutwyler,
Ann. Phys. \textbf{158}, 142 (1984).

\bibitem{Bilenky69}
S.\:M.~Bilenky, Nguyen Van Hieu, L.\:L.~Nemenov and F.\:G.~Tkebuchava,
Yad. Fiz. \textbf{10}, 812 (1969);
Sov. J. Nucl. Phys. \textbf{10}, 469 (1969).

\bibitem{densmatr04}
O.~Voskresenskaya,
J. Phys.~B: At. Mol. Opt. Phys. \textbf{36}, 3293 (2003);\\
L.~Afanasyev, C.~Santamarina, A.~Tarasov and O.~Voskresenskaya,
J. Phys.~B: At. Mol. Opt. Phys. \textbf{37}, 4749 (2004);
arXiv:hep-physics/0407110.

\bibitem{linalg}
G.\:H.~Golub and C.~Van Loan,
\textit{Matrix computations};
3rd ed.,~---
The Johns Hopkins University Press, 1996.

\bibitem{LAPACK}
E.~Anderson \textit{et al.},
\textit{LAPACK Users' Guide}, 3rd ed.,
Society for Industrial and Applied Mathematics,
Philadelphia, PA, 1999,
ISBN 0-89871-447-8.

\bibitem{DIRAC2005}
B.~Adeva \textit{et al.} (DIRAC Collaboration),
Phys. Lett.~B \textbf{619}, 50 (2005).

\end{thebibliography}
\end{document}